\documentclass[letterpaper]{article} 
\usepackage{aaai2026}  
\usepackage{times}  
\usepackage{helvet}  
\usepackage{courier}  
\usepackage[hyphens]{url}  
\usepackage{graphicx} 
\usepackage{natbib}  
\usepackage{caption} 
\usepackage{algorithm}
\usepackage{listings}
\usepackage{algpseudocode}

\usepackage{newfloat}
\usepackage{listings}
\DeclareCaptionStyle{ruled}{labelfont=normalfont,labelsep=colon,strut=off} 
\floatstyle{ruled}
\newfloat{listing}{tb}{lst}{}
\floatname{listing}{Listing}
\title{AI for Distributed Systems Design:\\Scalable Cloud Optimization Through Repeated LLMs Sampling And Simulators}
\author {
    Jacopo Tagliabue
}
\affiliations{
    Bauplan Labs\\
    New York City, NY USA\\
    jacopo.tagliabue@bauplanlabs.com
}

\begin{document}

\maketitle

\begin{abstract}
We explore AI-driven distributed-systems policy design by combining stochastic code generation from large language models (LLMs) with deterministic verification in a domain-specific simulator. Using a Function-as-a-Service runtime (\texttt{Bauplan}) and its open-source simulator (\texttt{Eudoxia}) as a case study, we frame scheduler design as an iterative \emph{generate--and--verify} loop: an LLM proposes a Python policy, the simulator evaluates it on standardized traces, and structured feedback steers subsequent generations. This setup preserves interpretability while enabling targeted search over a large design space. We detail the system architecture and report preliminary results on throughput improvements across multiple models. Beyond early gains, we discuss the limits of the current setup and outline next steps; in particular, we conjecture that AI will be crucial for scaling this methodology by helping to bootstrap new simulators.
\end{abstract}

\begin{links}
    \link{Simulator}{https://github.com/BauplanLabs/eudoxia}
    \link{Code}{https://github.com/BauplanLabs/AI-for-Distributed-System-Design}
\end{links}\section{Introduction}

Distributed cloud systems have many degrees of freedom: what is the optimal configuration for a Kubernetes cluster? What is the best caching policy given the history of read access to a system? In the majority of cases, experts turn to manually crafted rules \cite{s20164621}—i.e., programs whose input is the system state at a given \(t\) (and possibly historical traces) and whose output is the recommended infrastructure change for \(t+1\).

Rules-as-programs are interpretable and easy to run, but they are costly to scale, especially in business-to-business (B2B) settings. On the one hand, deep domain expertise is needed to craft sensible heuristics; on the other hand, even a ``perfect program'' crafted for customer \(k\) will not necessarily generalize to \(k+1\). While rules may have degrees of freedom themselves, even when heuristics or machine learning \cite{9659513,Ling2024LAVALV} are used to set parameters, we are still searching for the optimal settings for \(k+1\) within the same space as \(k\).

The rapidly improving coding capabilities of Large Language Models (LLMs) suggest a radically different approach: instead of manually crafting distributed-system policies, we could automatically \textit{generate} and \textit{evolve} them; as the cost of producing policies approaches zero, the search for optimal rules for customers \(k\) and \(k+1\) can now occur in different spaces altogether, as defined by semantically different programs.

In \textit{this} paper, we share preliminary results in applying LLMs to a concrete distributed-systems challenge. In particular, we summarize our contributions as follows:

\begin{itemize}
    \item We motivate our investigation with a real-world challenge, i.e., improving scheduling in a Function-as-a-Service (FaaS) runtime (\texttt{Bauplan}~\cite{bauplan2024}) through an existing open-source simulator (\texttt{Eudoxia}~\cite{eudoxia2024}).
    \item We pair a deterministic system simulator with the code-generation and reasoning abilities of frontier AI models. While simulators have been historically designed for \textit{humans} to experiment without the complexity of real-world deployments \cite{4488918}, we repurpose them as \textit{verifiers} for machine-generated ideas, a pattern that echoes recent successes in neuro-symbolic theorem proving \cite{jiang2023draftsketchproveguiding}.
    \item We release open-source code with a minimal, easy-to-extend reference implementation of the generate-and-verify loop.\footnote{Code will be made available with the camera-ready version.} Our prototype shows significant improvements in throughput over baseline policies across multiple frontier models.
\end{itemize}

The paper is organized as follows: Section~\ref{sec:industry} gives a brief overview of the target system and the available simulator; we then describe the sampling loop in Section~\ref{sec:discovery} and present our preliminary experiments in Section~\ref{sec:experiments}. We conclude by describing our roadmap and drawing general conclusions on the virtuous circle between code generation and simulations.

While the present implementation is tied to a specific cloud environment, scheduling is a general challenge, and our neuro-symbolic methodology is arguably scalable across the system landscape. For these reasons, we believe our work to be interesting to a broad range of practitioners, as it sits at the intersection of reasoning-as-code in LLMs, testing in distributed systems, and practical industry challenges.

\section{Industry Background}
\label{sec:industry}

To make the work self-contained, we provide a brief description of the salient design decisions behind the target FaaS runtime and a quick introduction to the FaaS simulator that will be used as the \textit{verifier} in our experiments. We refer the interested reader to the relevant literature for a deeper understanding of both \texttt{Bauplan} \cite{Tagliabue2023BuildingAS,cdms2025,10.1145/3650203.3663335,10825377,tagliabue2025safeuntrustedproofcarryingai} and \texttt{Eudoxia} \cite{eudoxia2024}.

\subsection{A FaaS for Data: Bauplan Design}

\texttt{Bauplan} is a composable \cite{10.14778/3603581.3603604} data lakehouse \cite{Zaharia2021LakehouseAN}, characterized by a unified execution model for synchronous and asynchronous data workloads. Departing from traditional architectures with heterogeneous abstractions, \texttt{Bauplan} adopts a pure FaaS model for all data operations \cite{bauplan2024}: both SQL queries and multi-language data pipelines are decomposed into directed acyclic graphs (DAGs) of functions.

\texttt{Bauplan}'s cloud architecture follows the standard separation of one global \textit{control plane} and per-customer \textit{data planes}. The \textit{scheduler}, sitting in the control plane, receives user requests through a public API; workers in a data plane are off-the-shelf VMs, providing stateless compute capacity to run the desired functions over data in object storage. While the FaaS execution model provides elegant abstractions for both end users and system developers alike, it also places extraordinary demands on the scheduler, which must efficiently pack functions onto available resources while respecting dependencies and latency requirements -- e.g., an interactive query should preempt a long-running job as users expect interactivity from the former, but not the latter. The challenge is particularly complex in such a system due to the interplay of two dimensions of variation:

\begin{itemize}
    \item \textbf{Within-customer diversity:} within a data plane, workloads vary widely in \textit{time}—from sub-second interactive queries to multi-hour batch pipelines—and in \textit{scale}—from dozens to hundreds of millions of rows.
    \item \textbf{Across-customer diversity:} depending on their data estate and concurrency requirements, two customers may differ greatly with regard to available resources and the workloads they run.
\end{itemize}

\texttt{Bauplan} is therefore an ideal testbed for exploring AI-driven optimizations—the B2B complexity justifies going beyond one-size-fits-all scheduling policies, but the problem is easy enough to state and constrained enough to permit meaningful evaluation.

\subsection{A FaaS Simulator: Eudoxia Design}

\begin{figure}
    \centering
    \includegraphics[width=\columnwidth]{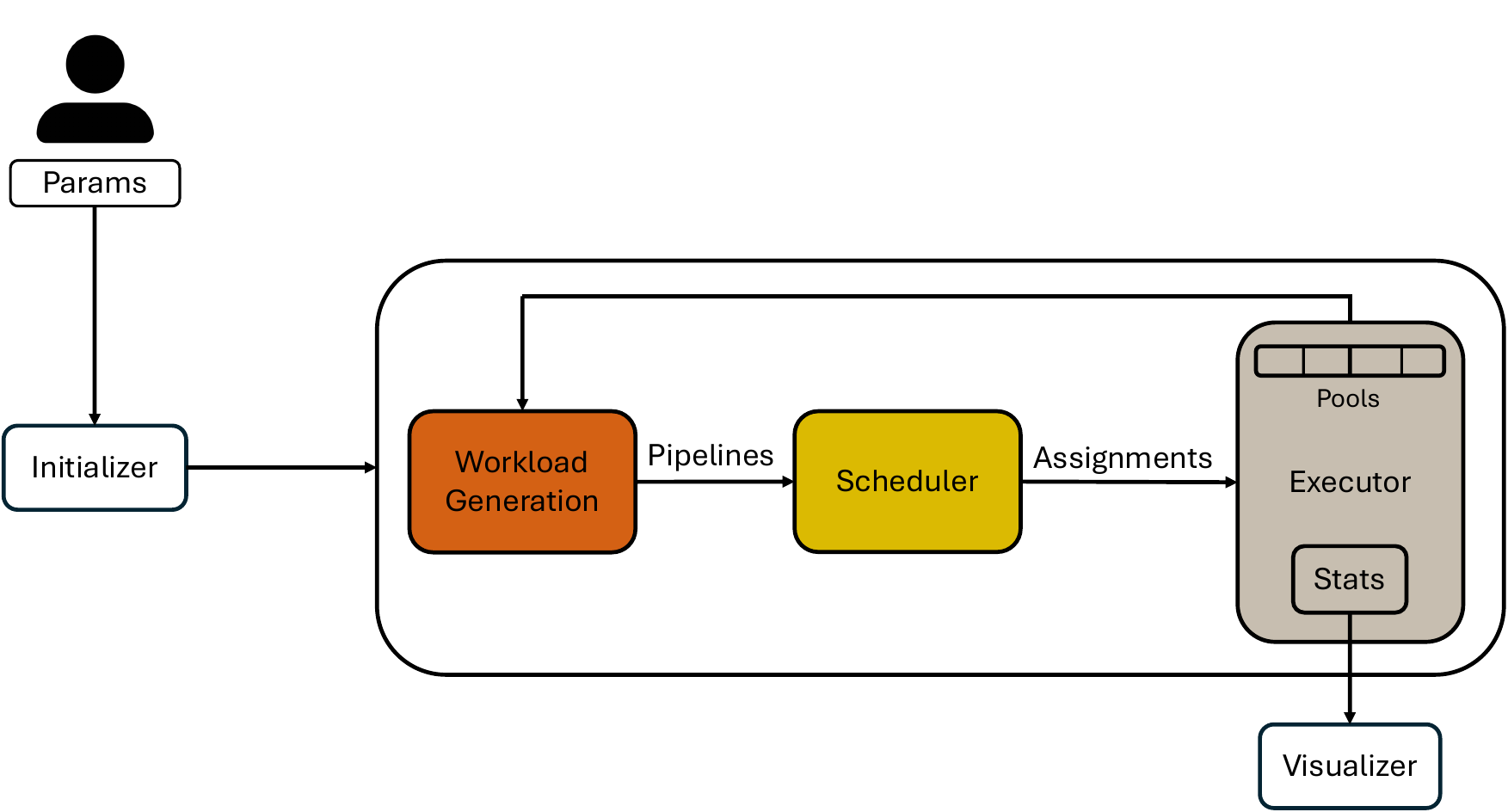}
    \caption{Simulator architecture. Users set parameters and pass them to the initializer, which starts a loop of three components—the Workload Generator, Scheduler, and Executor. Once that loop completes, visualizers or other downstream applications can access execution statistics.}
    \label{fig:arch}
\end{figure}

Recognizing the challenge of iterating on distributed systems \cite{daodistributed}, \texttt{Eudoxia} was developed to enable rapid prototyping and evaluation of scheduling policies before investing in real-world changes. The simulator provides a deterministic execution environment that models key aspects of FaaS scheduling: function arrival, resource allocation, execution-time modeling, and completion tracking. Figure~\ref{fig:arch} presents a high-level overview of the architecture, with the three main modules: Workload Generator, Scheduler, and Executor.

For our purposes, we can consider the Generator and Executor as providing the black-box scaffolding needed to instantiate and evaluate \textit{any} scheduling policy that gets \textit{registered} in the Scheduler. As a stripped-down example\footnote{Full code is available in the \texttt{Eudoxia} open-source repository.}, the following listing shows the required functions for a baseline FIFO policy:

\begin{lstlisting}[showstringspaces=false,columns=fullflexible,language=Python,numbers=none,caption=A FIFO baseline policy]
@register_scheduler_init(key='naive')
def naive_pipeline_init(s):
    s.waiting_queue = []

@register_scheduler(key='naive')
def naive_pipeline(s, failures, pipelines):
    # populate the queue
    for p in pipelines:
        ops = [op for op in p.values]
        # queue logic here

    suspensions = []
    assignments = []
    for pool_id in range(s.executor.num_pools):
        # FIFO logic here

    return suspensions, assignments
\end{lstlisting}

Users can implement new scheduling algorithms by matching the required type signature: the function accepts a set of pipelines from the Generator and returns a list of new allocations and preemptions to the Executor. At the end of a simulation, the system returns metrics of interest (e.g., throughput, \(p_{99}\) latency, and the number of failed DAGs).

Crucially, the simplicity of the interface makes it amenable to LLM generation, while at the same time the code-first approach and built-in APIs to simulate resource pools, priority queues, and hardware consumption make it possible to explore a potentially large and sophisticated design space.

\section{AI-Driven Policy Discovery}
\label{sec:discovery}

Frontier models’ ability to answer complex questions can be effectively augmented without access to their weights by repeated sampling at inference time \cite{Brown2024LargeLM}. Our approach similarly treats code generation -- i.e., a computational policy compliant with the \texttt{Eudoxia} interface and APIs -- as a sampling process from a black-box model, with context modified between inferences. As such, the system operates as an iterative refinement loop, where each iteration generates a new policy, evaluates it, and incorporates the results to guide subsequent generation.

\subsection{System Architecture}

Built on a black-box LLM (accessed via LiteLLM\footnote{\url{https://docs.litellm.ai/}}), the system comprises three modules:

\begin{enumerate}
    \item the \textbf{policy generator} takes as input the current context and produces a new Python policy algorithm (via an LLM call, syntax checks, and string manipulation);
    \item the deterministic \textbf{simulator} takes as input a generated Python policy and runs it on standardized traces, producing key performance metrics (e.g., throughput, \(p_{99}\) latency) and (optionally) verifying safety invariants via runtime monitors (e.g., no oversubscription, bounded waiting);
    \item the \textbf{context manager} takes as input the current context and the result of a simulation run, and produces an updated context (for the next iteration) that includes a summary of the results obtained by the run and the given policy—for example, early policies typically have an imprecise understanding of the semantics of \texttt{Eudoxia} APIs, so reporting precise failures back to the model helps it get unstuck.
\end{enumerate}

As the interfaces are well-defined, it is important to note that in our design all four components (the three modules and the LLM) can be changed and improved independently: not only can better LLMs be trivially tested with a simple configuration change, but the system will also immediately benefit from any independent improvement in the simulator software itself.

\subsection{The Discovery Loop}

\begin{algorithm}[tb]
\caption{Sampling Loop for Policy Discovery}
\label{alg:algorithm}
\textbf{Input}: System prompt, user prompt\\
\textbf{Parameters}: LLM model, target metric, workload parameters, number of iterations, simulation parameters (e.g., number of CPUs, number of pools)\\
\textbf{Output}: Best policy with score (e.g., latency, throughput)
\begin{algorithmic}[1]
\State Initialize context with system prompt and user prompt
\For{iteration $i$ in $1$ to $K$}
    \State Request new policy from LLM given context
    \State Parse and validate generated Python code
    \If{code is valid}
        \State Run policy in simulator with test workload
        \State Collect performance metrics and trace data
        \State Synthesize structured feedback
        \State Append (policy, metrics, feedback) to context
    \Else
        \State Append (policy, error, suggestions) to context
    \EndIf
    \If{context exceeds token limit}
        \State Compress context
    \EndIf
\EndFor
\State Return best policy with score
\end{algorithmic}
\end{algorithm}

The pseudo-code in Algorithm~\ref{alg:algorithm} describes the essential steps in the sampling loop. Starting from a context built from a \textit{system prompt} and a \textit{user prompt} (see below), workload and simulation parameters define a standard, deterministic simulation environment in \texttt{Eudoxia} that persists for the entire experiment. The loop itself is straightforward: the policy generator takes the current context and produces a scheduling policy as Python code; if the code is correct, the simulator runs it against the environment and records the target metric over time; finally, the manager evaluates the results and updates the LLM context for the next iteration with a summary of what happened in the current turn.

This iterative process allows the LLM to learn from both successes and failures. Syntax errors teach it about interface constraints, while performance feedback guides it toward more effective strategies. Crucially, because the generated policies are readable Python code, human operators can inspect, understand, and modify them—as part of the loop, all policies are logged as executable Python files for asynchronous debugging and human evaluation.

\subsection{Context Management}

At the start of an experiment, the context comprises only the \textit{system} and \textit{user} prompts. The system prompt provides a high-level view of the FaaS system, the simulator, and key metrics. Then, in a few-shot fashion, existing policies are provided as examples, with an emphasis on respecting the interfaces. The user prompt asks the model to generate a new policy starting from the baseline policy. The prompt also explicitly states which metric to optimize.

As the experiment progresses, the LLM builds a larger context containing previous policies and their results: rather than generating entirely new approaches each time, the context manager nudges the LLM to analyze previous attempts and propose targeted improvements.
 
\section{Experiments}
\label{sec:experiments}

We present preliminary results from running the discovery loop with frontier models to improve the \textit{throughput} of the FIFO policy that comes with \texttt{Eudoxia}. FIFO is a good choice for our present feasibility study because it is easy to inspect even for programmers not familiar with \texttt{Bauplan} architecture, but it is still a working policy that can be used as the starting point for code evolution.

The setup is as follows. We first generate six realistic traces using \texttt{Eudoxia}’s built-in APIs: three sets of two traces each are generated by specifying different parameters for the random distribution, and then varying the seed to get two out of the same parameter set. For each LLM, we run the sampling procedure for \(50\) iterations and collect the best-performing policy by throughput (i.e., median throughput across traces). Finally, we report the improvement relative to the baseline, the total cost (as reported by LiteLLM), and the total time for the full loop.
 
Table~\ref{tbl:exp} summarizes the results of our experiments, comparing Sonnet (with temperature \(0.7\)), Opus (with temperature  \(0.7\)), GPT5 and GPT5-mini (both with \textit{high} reasoning effort).\footnote{We report the best policy results out of three independent runs for each model. All experiments were performed between October 19--21, 2025.} Interestingly enough, models vary wildly in their ability to provide good policies, with GPT5 providing the biggest improvement and taking the longest. The failure of GPT5-mini to produce \textit{any} improvement at all over three trials shows that the baseline is indeed non-trivial even for capable models.

\begin{figure}
    \centering
    \includegraphics[width=0.8\columnwidth]{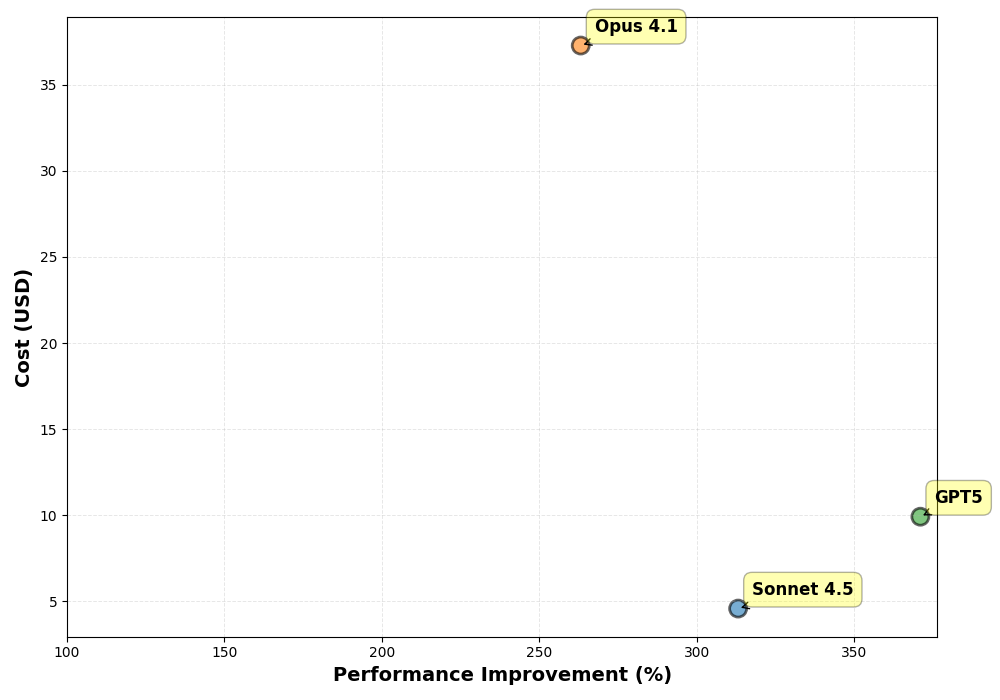}
    \caption{Price-performance landscape. \textit{X-axis}: LLM performance as measured by improvement over the baseline for the best generated policy (omitting models with no improvements). \textit{Y-axis}: cost in USD.}
    \label{fig:results}
\end{figure}

Figure~\ref{fig:results} provides a visual representation of how models compare along the price-performance dimensions. For the purpose of this preliminary investigation, we are concerned about the general feasibility of the proposed methodology and the questions raised for the future of co-designing distributed systems with AI. Nor the setup, nor the magnitude of the improvement over FIFO is intended to convey a judgment on model strength or generalizable improvement, but only to illustrate how even a single change in the loop may result in drastically different outcomes.

\begin{table}
\caption{Policy improvements across models (best in bold)}
\begin{center}
\begin{tabular}{|c|c|c|c|}
\hline
\textbf{\textit{Model}} & \textbf{\textit{Improv.}} & \textbf{\textit{Cost (USD)}} & \textbf{\textit{Total time (s)}} \\
\hline
Sonnet 4.5 & 313.2\% & 4.587 & 2785 \\
Opus 4.1 & 263.2\% & 37.27 & 2158\\
\textbf{GPT5} & \textbf{371.1\%} & 9.92 & 8292 \\
GPT5-mini & 0\% & 1.65 & 7669 \\
\hline
\end{tabular}
\label{tbl:exp}
\end{center}
\end{table}

\section{Conclusion and Future Work}

This work represents an early exploration into AI-driven systems optimization. While LLM inference is rapidly being commoditized, our project was enabled by an independently developed simulator for the target system. As we defer thorough investigation of this point to future work, we cannot help but wonder whether LLMs could help bootstrap the verifier itself, enabling this methodology to scale without relying solely on manually built simulators -- for example, it has been conjectured that predicting system-level outcomes from specifications could ``open new avenues for creating universal simulators for complex systems.'' \cite{akhauri2025performancepredictionlargesystems}

We conclude by highlighting a few directions for future work, along the key dimensions of robustness (``how much can we trust the results?''), accuracy (``how do we get better policies?'') and extensibility (``how do we extend the methodology to a different system?''):

\begin{itemize}
    \item \textbf{Robustness}: sampled policies will match real-world outcomes only if the simulator is \textit{representative enough} -- a nontrivial amount of engineering and statistical inference is required to build more confidence in the verifier before scaling up costly experiments.
    \item \textbf{Accuracy}: immediate prompt-related optimizations are prompt evolution \cite{agrawal2025gepareflectivepromptevolution} and better context management, e.g., \texttt{Eudoxia} produces fine-grained logs which arguably contain useful details. Furthermore, recent successes with evolutionary computation suggest a way to perform a structured exploration of the space \cite{nagda2025reinforcedgenerationcombinatorialstructures}; finally, training smaller models with RL \cite{deepseekai2025deepseekr1incentivizingreasoningcapability} could yield policies that are both effective and efficient. 
    \item \textbf{Extensibility}: \texttt{Bauplan} is a good initial case study because it is complex enough to be interesting, but also simple enough that current LLMs can realistically make progress. Starting from our initial success with FaaS-for-data, we could try and adapt the simulator to model more general serverless systems, e.g., a multi-tenant, Lambda-like FaaS -- it is an open question how far sampling can get us when modeling more complex and concurrent distributed systems.
\end{itemize}

The convergence of AI and systems optimization represents a fundamental shift in how we approach complex engineering challenges. Rather than carefully crafting algorithms by hand, we can leverage AI to explore vast design spaces while maintaining human interpretability \cite{cheng2025barbariansgateaiupending}. While our experiments represent only a first prototype, they offer a glimpse of a new dawn, through the practical lens of a real-world FaaS. 

\section{Acknowledgments}
The author wishes to thank Tapan Srivastava and Tyler Caraza-Harter for their work on \texttt{Eudoxia}, and the \texttt{Bauplan} team for the extraordinary work in building a FaaS runtime from scratch. Thanks to Ciro Greco and Federico Bianchi for helpful comments on a preliminary version of this draft.

\bibliography{aaai2026}

\end{document}